\documentclass[12pt]{IEEEtran}
\IEEEoverridecommandlockouts
\usepackage{cite}
\usepackage{amsmath,amssymb,amsfonts}
\usepackage{bbm}
\usepackage{algorithm}
\usepackage{algorithmic}
\usepackage{graphicx}
\usepackage{caption}
\usepackage{textcomp}
\usepackage{xcolor}
\usepackage{url}
\usepackage{booktabs}
\usepackage{siunitx}
\usepackage{comment}

\usepackage{multirow}
\def\BibTeX{{\rm B\kern-.05em{\sc i\kern-.025em b}\kern-.08em
    T\kern-.1667em\lower.7ex\hbox{E}\kern-.125emX}}
\begin{document}

\title{Federated Dynamic Spectrum Access
}
\author{\IEEEauthorblockN{Yifei Song, Hao-Hsuan Chang, Zhou Zhou, Shashank Jere and Lingjia Liu}
\thanks{The authors are with the Department of Electrical and Computer Engineering, Virginia Polytechnic Institute and State University, Blacksburg, VA, 24061}
\thanks{The corresponding author is L. Liu (ljliu@ieee.org).}
}

\maketitle

\begin{abstract}

Due to the growing volume of data traffic produced by the surge of Internet of Things (IoT) devices, the demand for radio spectrum resources is approaching their limitation defined by Federal Communications Commission (FCC). To this end, Dynamic Spectrum Access (DSA) is considered as a promising technology to handle this spectrum scarcity. However, standard DSA techniques often rely on analytical modeling wireless networks, making its application intractable in under-measured network environments. Therefore, utilizing neural networks to approximate the network dynamics is an alternative approach. In this article, we introduce a Federated Learning (FL) based framework for the task of DSA, where FL is a distributive machine learning framework that can reserve the privacy of network terminals under heterogeneous data distributions. We discuss the opportunities, challenges, and opening problems of this framework. To evaluate its feasibility, we implement a Multi-Agent Reinforcement Learning (MARL)-based FL as a realization associated with its initial evaluation results.

\end{abstract}

\begin{IEEEkeywords}
5G, Federated Learning (FL), Dynamic Spectrum Access (DSA), Dynamic Spectrum Sharing (DSS), Multi-agent Learning, and Deep Reinforcement Learning
\end{IEEEkeywords}

\section{Introduction}

Due to the unrolling of the Fifth-Generation cellular networks (5G), the number of wireless devices is expected to arrive at the scale of trillions in the next decade. 5G aims to extend its coverage, transmission bandwidth, and reduce communications latency, paving the way to emerging applications, such as smart home, autonomous vehicles, unmanned aerial vehicle and augmented reality. According to Cisco's Annual Internet Report, Internet of Things (IoT) as a major developing category, there will be $5.3$ billion internet subscribers and $29.3$ billion devices connecting to the network by $2023$. Meanwhile, the average of the 5G speed is expected to reach $575$ Mbps, which is 13 times higher than average mobile connection~\cite{cisco2020cisco}. 
In 2021, Federal Communications Commission (FCC) announces bidding over $81$ billion dollars for a total of $280$ MHz licenses of usage for the aforementioned frequency band. To efficiently use the scarce and expensive spectrum resources, 
the Third Generation Partnership Project (3GPP) has been working on standardization to enable 5G New Radio (NR) service in unlicensed spectrum as part of the LTE Licensed Assisted Access initiative (LAA)~\cite{5GDSS_2019}.

Conventional spectrum management policy assigns frequency bands to wireless operators for exclusive use.
%
%
%
Dynamic Spectrum Sharing (DSS) is a promising technology for 5G and Beyond 5G networks to improve spectrum efficiency.
%
DSS adopts a hierarchical spectrum access structure that allows Secondary Users (SUs), unlicensed users, to access the spectrum resource without causing substantial interference to primary licensed users. 
Therefore, the spectrum utilization ratio can be increased by DSS significantly.

The existing DSS strategies can be divided into learning based and non- learning based methods.
Most non-learning based DSS appoaches need accurate network information such as channel state information and user behavior to manage the spectrum resource allocation.
Since the wireless environment is highly dynamic, tracking the accurate network information requires frequent communications between SUs and Primary User (PUs), which imposes significant control overhead in wireless network operations, especially when the number of users is large.
Furthermore, developing a solvable mathematical model corresponding to the highly complex wireless network operations is extremely challenging or not possible at all.
To address these issues, machine learning (ML) based methods have been applied to the DSS problem by learning from the data in the wireless environment~\cite{DSM}.
Federated learning (FL) is a hybrid method that takes advantage of both centralized learning and distributed learning.
FL executes model training in a decentralized way and preserves data where it is created and no crude data are transferred to mitigate the communication overhead, protect user privacy and achieve the desired system performance.
FL is introduced in the 3GPP Technical Report (TR) 22.874 as a potential candidate to support ML model transfer and distribution in the 5G system.

In this article, we provide the background knowledge of the DSS system and introduce different approaches for solving the spectrum access problem in the DSS system.
Meanwhile, we describe the shortage of current network and potential challenges with the rapid 5G technology and IoT development.
Then, we discuss the outstanding features of federated learning that distinguish itself from the traditional ML framework.
To provide a concrete example, we apply the FL to solve the DSA problem in the Citizens Broadband Radio Service (CBRS) system, and we evaluate the performance of the FL by comparing it with other learning methods.
Finally, the article concludes the future research challenges of applying FL to different applications. 


\section{PRELIMINARIES AND OVERVIEW}
\label{Preliminaries_and_overview}

\subsection{Dynamic Spectrum Sharing}

To efficiently utilize spectrum resources, two types of spectrum management mechanisms are utilized: static and dynamic.
Static spectrum sharing groups and reorders all spectrum resources to assign the same portion back to Service Providers (SPs). 
The licensed SPs schedule these spectrum resources to their subscribers accordingly.%
In dynamic spectrum sharing (DSS), spectrum resources are dynamically allocated to both licensed SPs and unlicensed SPs to 
provide an efficient way to utilize available radio resources and alleviate the spectrum shortage without adding new spectrum resources for unlicensed SPs. 
A special DSS strategy is the opportunistic DSS system where unlicensed users are allowed to access the radio resources when licensed users are idle.
Licensed users and unlicensed users are defined as Primary Users (PUs) and secondary users (SUs), respectively.
Accordingly, the distribution of nearest spectrum holes can be determined by the marked base station tailored for corresponding access strategies.~\cite{chen2018qos}
\subsection{Citizen’s Broadband Radio Service (CBRS) System}

The CBRS system has been made available for shared connectivity under a three-tiered spectrum access model: Incumbent User (IU), military uses and satellite stations, have the highest priority~\cite{CBRS}. To make sure they are shielded from users in the lower tiers. The spectrum is then licensed to commercial customers who purchase Priority Access Licenses (PAL) for a certain area and periods of time, with each PAL having a bandwidth of 10 MHz and the CBRS scheme leasing up to 70 MHz as PAL. General Authorized Access (GAA) tiers has the lowest priority stands for unlicensed users which provides GAA subscribers with at least 80 MHz of spectrum. Spectrum Access System (SAS) is used in charge of spectrum sharing management which guarantees the bandwidth is sufficient for Incumbent User (IU). Meanwhile, if all IUs and PAL users are idle, unused PAL bandwidth would be provided to GAA users, allowing GAA users to access the whole 150 MHz spectrum. Remark that all PAL and GAA users should cease using the spectrum if the IUs need it. The SAS does not provide interference protection to GAA users, so multiple GAA users accessing the same channel at the same time can cause interference. For simplicity purposes in this work, we call IUs and PALs are both Primary Users (PU) and GAA users as Secondary Users (SU), where the devices operate on a 150 MHz bandwidth between 3.55 and 3.7 GHz. Note that SAS makes SUs channel access decisions. In other words, SUs 
make no attempt to sense PUs activities but ask permision from SAS.

\subsection{Multi-Agent Reinforcement Learning (MARL)}
ML has been widely used in wireless communications including DSS systems where ML algorithms are used to allocate spectrum resources more effectively. 
Different ML approaches are discussed in~\cite{RubayetAI} to provide automated and efficient DSS algorithms. 
As discussed in~\cite{RubayetAI}, Deep Reinforcement Learning (DRL) is a natural tool for dynamic spectrum access and sharing.
Specifically, the DRL agent takes an action based on an observation of the environment state, receives a reward from the environment that evaluates the taken action, and transits into a new state.
The goal of the DRL agent is to find a policy that optimizes the cumulative reward.
In MARL, more than one agent are involved in the system which turns into an optimization problem facing all agents' policies. This is because individual rewards received by each agent and the system state are affected by other agents' policies~\cite{bucsoniu2010multi}. 
MARL allows agents to communicate with each other and parallel processing once agents receive distributed tasks. For example, agents can share their experiences with others to boost their learning convergence. Furthermore, MARL system is able to add new agents into the system and replace inactive agents.

However, MARL always suffers from unbearable computation time, where the complexity follows an exponential rule to the problem's dimensionality, environment's nonstationarity, and the configurations of exploration and exploitation trade-off~\cite{bucsoniu2010multi}.

\subsection{Federated Learning (FL)}
\begin{table*}[h!]
  \begin{center}
    \caption{Comparison of Different ML Approaches.}
    \label{tab:table1}
    \begin{tabular}{|l|l|l|}
    \toprule
        \textbf{ML type} & \textbf{Pros}   & \textbf{Cons} \\ 
        \hline
        \multirow{3}{*}{Centralized Learning}& High Accuracy & Slow Learning Rate\\ 
        
        & Users Cooperation & Large Communication Overhead\\ 
        & Large Data Base &Data Privacy Risk\\
                \hline
        \multirow{3}{*}{Distributed Learning} & High Learning Rate & No User Feedback\\
        &Low Communication Overhead & Limited Data Base\\
        &Data Privacy Risk &Worse System Performance\\
        \hline
        \multirow{2}{*}{Parallel Learning} & No Communication Constraints & i.i.d Data \\
        &Efficient Execution & No Device Cooperation\\
        \hline
        \multirow{4}{*}{Federated Learning} & Can Handle Non-i.i.d Data & Slow Convergence\\
        &Limited User Feedback &Communication Constrains\\
        &Data Security &High Training Variance \\
        &Adaptable Reactions to Dynamic Environment & \\
        \hline
        \bottomrule
    \end{tabular}
  \end{center}
\end{table*}

\begin{table*}
\centering
\caption{FL Convergence Analysis.}
\begin{tabular}{ |p{2cm}||p{3cm}|p{3cm}|p{3cm}||p{3cm}|  } 
\toprule
\textbf{Method name} & \textbf{Model Quantization}   & \textbf{Partial Devices Participation}   & \textbf{Periodic Aggregation}     & \textbf{Data Heterogeneity}                        \\ 
\hline
FedAVG$-1$  &\checkmark     & \checkmark &\checkmark    & i.i.d\\
FedAVG$-2$  &    & \checkmark &\checkmark    & non-i.i.d\\
FedPAQ  &\checkmark   &\checkmark  &\checkmark    &i.i.d \\
FedPROX  &   &\checkmark  &\checkmark    &non-i.i.d \\
SCAFFOLD  &\checkmark   &\checkmark  &\checkmark    &non-i.i.d \\

\hline
\bottomrule
\end{tabular}
\label{simulation_parameters_table}
\end{table*} 
\subsubsection{Limitation of Conventional Approaches}

In conventional machine learning methods, user devices send their raw data to centralized server for training to create a model for interference in the future. Due to the massive number of data that must be used in the training process, the raw data contains private information. On the other hand, distributed learning only requires the cloud to send a model to users and trains the model with local users' dataset. No further communication is needed. However, in distributed learning, all users optimize their models without considering the variation of the environment due to limited data sharing and cooperation. 
\subsubsection{The Beauty of FL}
FL is a viable alternative for transitioning from a centralized learning to a more realistic decentralized paradigm to address the aforementioned concerns. FL integrates distributed training models product from local devices to create a global representation. In particular, each end user efficiently trains its local model from a limited dataset and sends model updates to a server aggregator. The server develops a global model based on collected information and distributes back to users. The global model accuracy can be assured by sufficient turns of communications. The Pros and Cons of traditional ML framework and FL can be found in table~\ref{tab:table1}.
\subsubsection{Optimization Analysis}
FL optimization leans on Stochastic Gradient Descent (SGD), which has been shown to be effective in training deep networks for statistical problems. To guarantee an unbiased estimate of the full gradient, independent and identically distributed (i.i.d) sampling of the training data is critical. In fact, it is impractical to assume that the observation of each device is i.i.d Furthermore, to achieve a high degree of communication reduction, model quantization, partial devices participation, and periodic aggregation are all needed to be considered. The author in~\cite{mohammadi2021differential} compares the most common approaches, FedAvg, FedPaq, FadProx and SCAFFOLD in table~\ref{simulation_parameters_table}. Related literature for FedAvg is either missing the quantization model compression or assumed i.i.d data. FedPaq is an ideal candidate in practice but the analysis is usually for i.i.d data type only. FadProx, a developed version of FedAvg, considers the non-i.i.d data types but omits the model quantization. Finally, it seems that SCAFFOLD satisfies all elements even author claims they achieve better performance compared with FedAvg. However, they have to double the communication expense since SCAFFOLD requires all devices to send an extra update of the control variable which is the same dimension as the model parameters.

\subsection{FL in Wireless Communication}
In wireless communication, federated learning distributes the model through the Broadcast Channel (BC), aggregates model updates via Multiple Access Channels (MAC). The aggregation process is usually the constraint in wireless communication systems due to limited uplink bandwidth and model updates are different from users which makes the process slow and expensive. In respect of improving communication efficiency, quantized information transmission, partial devices participation, and periodic aggregation are commonly used.
\subsubsection{Model Compression}
Even though FL collects only gradient parameters to alleviate the communication overhead, there are billions of devices~\cite{cisco2020cisco} in a wireless network that accumulates communication overhead. Therefore, instead of upload full-accuracy model parameters, local model updates can be compressed with low accuracy quantization operators. Overall, exchanging quantized updates reduces communication overhead.

\subsubsection{Partial Devices Participation}
Due to the heterogeneous data size in local devices and devices network, channel conditions can be dramatically different based on device location, transmit power, and noise level. It is not practical to have all devices participate in the whole training process. The huge number of participate devices results in an expensive communication process and communication overhead. Furthermore, devices may not able to respond( inactive), known as Straggler's effect. 

\subsubsection{Periodic Aggregation}
As with traditional distributed learning, aggregation in each training iteration leads in significant communication overhead. Thus, in FL, after the server distributes a model, participate devices perform a series of local iterations and periodically synchronize with the server to reduce communication overhead. It is important to note that the update direction can diverge from the global gradient direction, especially for non-i.i.d settings, because all devices do local updates in an unsynchronized manner.

It is worth noting that the above-discussed communication reduction tactics are inextricably coupled. The number of devices that can be active under a certain MAC channel capacity is closely tied to the payload of the model parameters. At the same time, with a constant MAC capacity restriction, there is a clear trade-off between the payload of the model parameters and the aggregation frequency.

\section{Configuration of MARL Enabled FL for DSA}

\begin{figure}[!t]
\includegraphics[width=1\linewidth,clip]{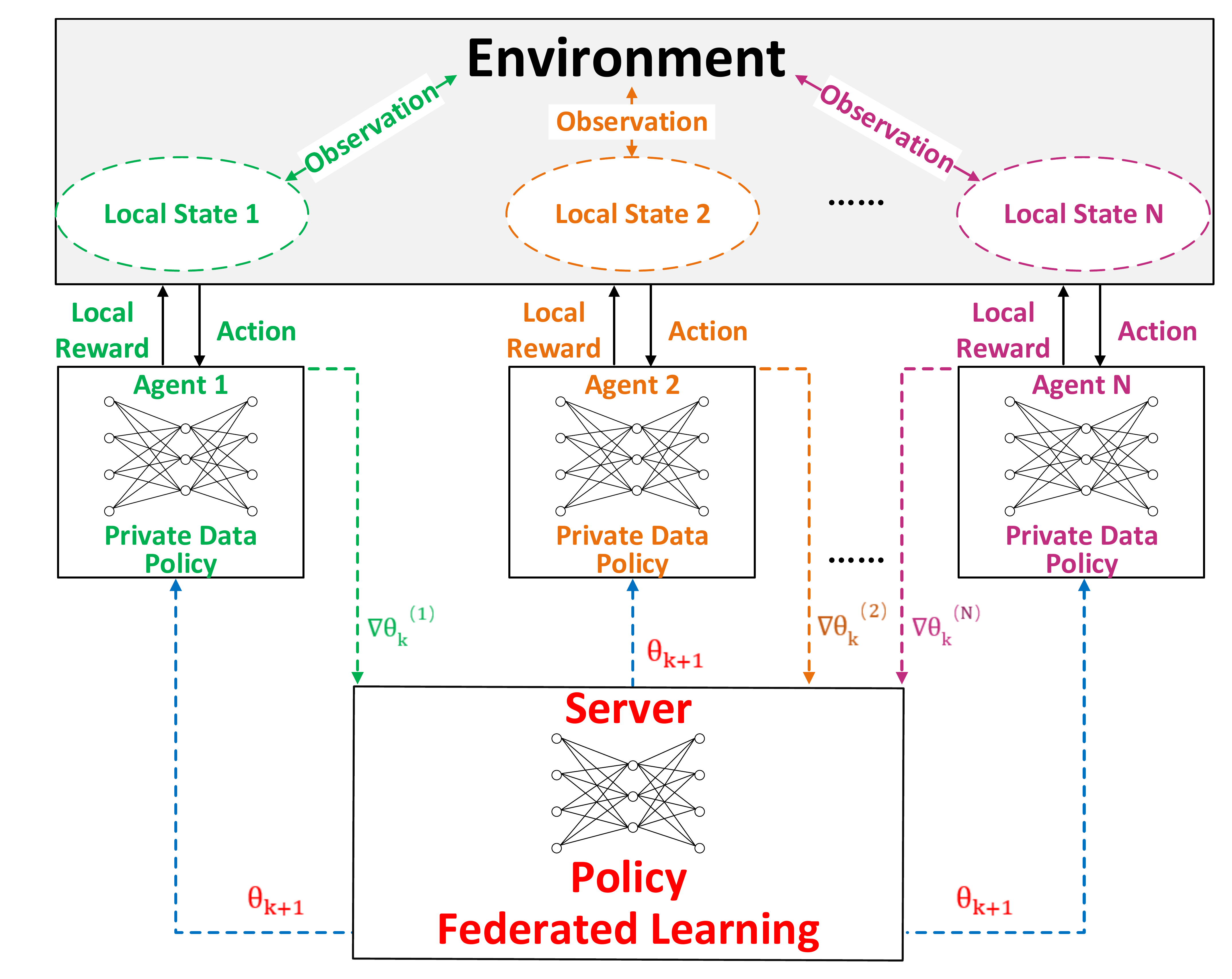}
\caption{Multi-agent reinforcement learning in federated learning framework.}
\label{FL_MARL}
\end{figure}
\subsection{System Framework}
\subsubsection{Outline}

Existing MARL algorithms assume that a joint reward is received by all agents, or each agent receives an individual reward but shares it with other agents.
However, this assumption may not be practical in some real-world applications because agents do not share their received observations and rewards for data privacy and security concerns.
In this MARL enabled FL system, each agent does not share its local observations and rewards with other agents, and updates its policy to maximize long-term local reward. MARL can be characterized as a pool of all agents' state space, observation space, action space, state transition probabilities, reward space, and discount factors. 
The objective of the system is to optimize the joint reward, which is the sum of the product of each agent reward and the weight of each agent. The outline of FL in MARL can be found in Figure~\ref{FL_MARL}. We select Signal-to-Interference-plus-Noise Ratio (SINR) as our quality measurement~\cite{DEQN} which characterises all the factors to SUs network, such as background noise, the transmission power, the interference between simultaneously transmission pairs. In addition, we choose channel capacity as the reward function. 
\subsubsection{Spectrum Access Policy}
We use a decentralized policy gradient method in our MARL system to optimize the joint reward. An initialized policy first distribute to all agents. The neural network at each agent is updated according to its own gradient. In each communication cycle, the agent empties its buffer, observes the environment, takes an action based on its policy, and receives a reward from the environment. After sufficiently repeating the above-mentioned steps, each agent obtains an updated local policy network. 
The updated local policy networks are then aggregated at a central server to renew the global policy network. In our framework, the policy network is implemented by a Recurrent Neural Network (RNN) with a softmax activation function in the output layer.
%

%
\subsubsection{User Selection}
\label{Q_learning}
To achieve an efficient wireless communication system, user selection is very important to accelerate convergence and counteract the non-i.i.d data bias. To this end, we consider conducting the user selection through a separate deep Q-learning framework. The agent learns to pick devices in each communication cycle to minimize system validation error with reduced communication overhead. Only model weights are conveyed in the proposed framework. Therefore, the agent does not need to gather or validate any data samples from mobile devices, retaining the same level of privacy in FL. Because there is an implicit relationship between the data distribution on a device and the local model weights acquired by running SGD on those data. The server exclusively relies on model weight information to identify which device may enhance the global model the most.

We model a SUs spectrum accessing strategy to utilize the spectrum resources efficiently in CBRS system. Assuming there are $N$ SUs and $M$ channels $(N>M)$, and each SU can only access one channel at a specific time. To avoid interference from unlicensed users, SUs are not allowed to transmit when PU is occupying the spectrum. 
However, each SU may interfere with each other.
To collaboratively avoid interference among SUs, we apply the previous defined spectrum access framework to SUs.
\subsection{Simulation Results}
\begin{figure}[!h]
    \centering
    \includegraphics[width=1\linewidth, height=0.75\linewidth]{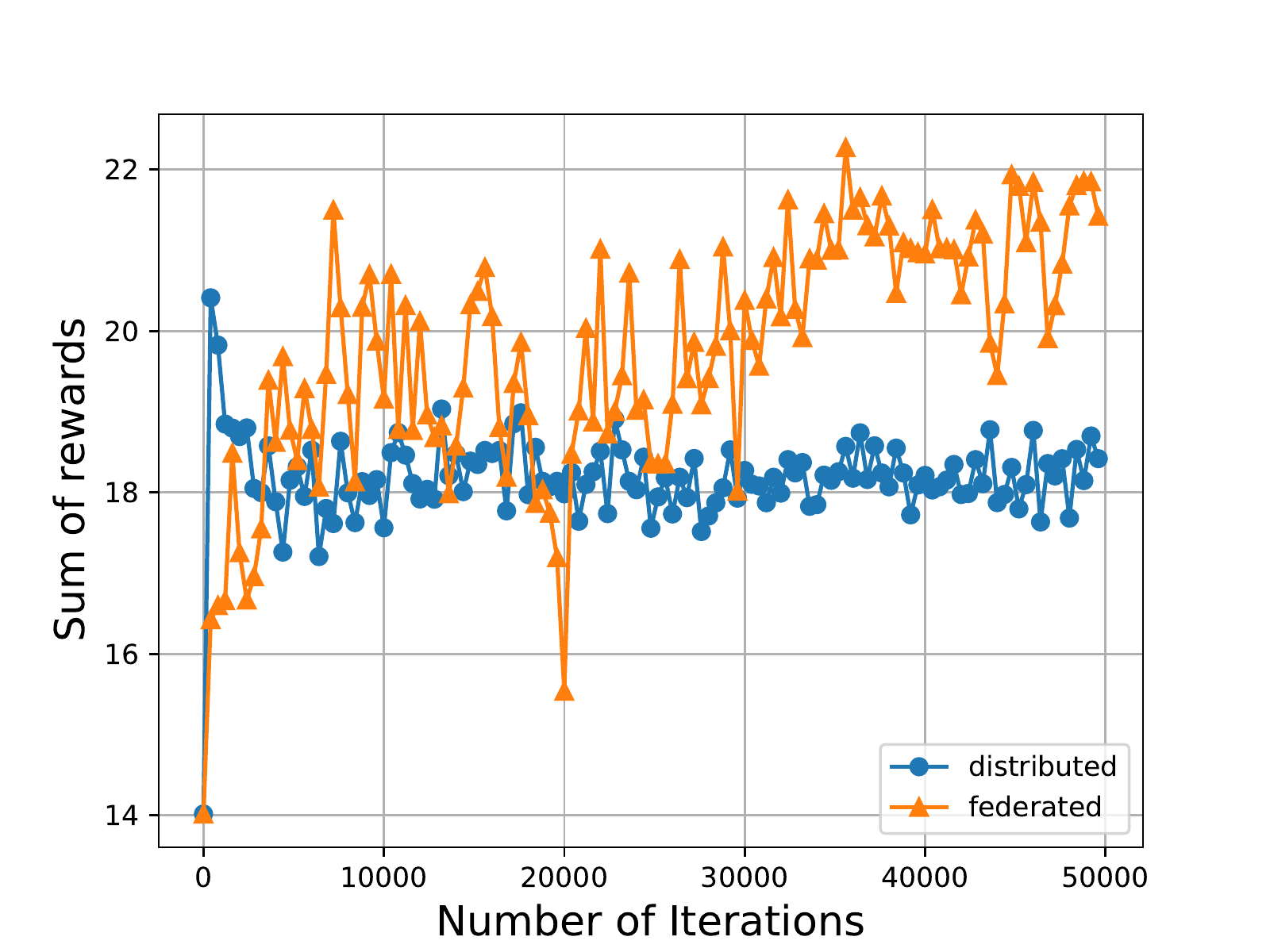}
    \caption{Federated Learning vs Distributed Learning.}
    \label{FL_time}
\end{figure}

\begin{figure}[!h]
    \centering
    \includegraphics[width=1\linewidth, height=0.75\linewidth]{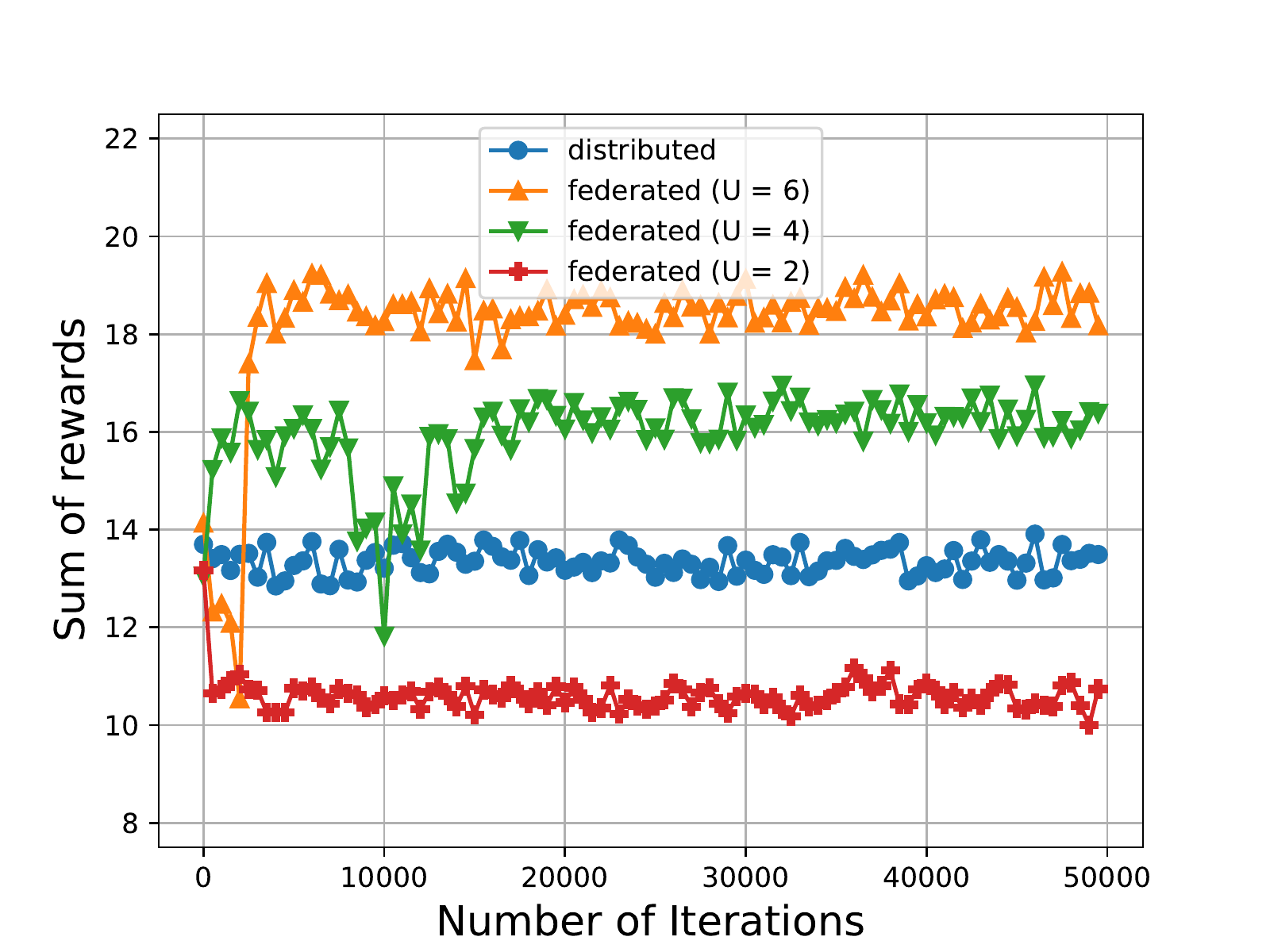}
    \caption{Sum rewards under participated users.}
    \label{Number_Users}
\end{figure}

We randomly generate transmitter and receiver pairs in a 400-meter by 400-meter area with 8 user pairs and 4 channels in the simulation set up. In the policy configuration, each agent chooses to access one of four channels or idle. In our simulation, we compare our MARL enabled FL with a conventional distributed learning approach. In the distributed learning setting, each user receives a training model from the cloud and starts to maximize their local rewards without any further communication. MARL enabled FL can provide better overall performance at the system level since it allows the cooperation among users through aggregating their policy networks.
Figure~\ref{FL_time} and Figure~\ref{Number_Users} indicate respected setting for our CBRS system. Figure~\ref{FL_time} shows that FL gives a better outcome than distributed learning, whereas the former one requires more communications. The result indicates a trade-off between communication overhead and achievable system throughput. Figure~\ref{Number_Users} investigates the federated learning framework under different number of participated users ~\ref{Q_learning}. It shows more selected users outcome higher sum of rewards. Meanwhile, the partial participation mechanism allows the system to be more flexible in choosing users with a better channel condition, energy constraints, and sufficient computation resources. 

\section{Research Directions and Challenges}

\subsection{Joint processing}
The aforementioned dynamic spectrum access strategy is only tailored to communication systems with a SAS like in the CBRS. Meanwhile, the RL observation purely relies on instantaneous power of other SUs. To further improve the spectrum efficiency, we can consider adding extended dimensions of the observation, such as the temporal and spatial statistics of the spectra.
Furthermore, the local agent can learn to predict users' access behaviors by adding a supplementary policy network based on their mobility and peak-time traffic. However, the implementation of the subsidiary neural network demands additional memory and energy resources. Aside CBRS systems, SUs are configured to sense the PUs' activities before their access. For instance, \cite{sensing} proposed combining energy detector, matched filter, and cyclic prefix to provide a less false alarm probability and high detection probability which can be further analyzed to strengthen our work.


\subsection{Privacy Issues} 
Maintaining maximum participated devices in the FL training process is always a problem especially in an irresponsible network environment. Furthermore, to save energy, battery-powered IoT devices incorporate strategies particular to distributed learning, opting out of rounds of training for a given job. The assurance of local user data privacy is also a standout feature of FL. Adversaries still have a potential to glean critical information from model changes. 
Although newer methods such as secure multiparty computation (SMC), differential privacy~\cite{McmahanFedML2017} or secure aggregation~\cite{bonawitz2017practical} seek to improve the privacy of federated learning, these approaches generally sacrifice inference performance in sake of privacy. Understanding and balancing these costs is a significant difficulty in implementing private federated learning systems, both theoretically and practically~\cite{FL_challenges_review}.

\subsection{Asynchronous FL Optimization} 
Although synchronous FL model provides better convergence guarantees, it is sensitive to the Straggler effect~\cite{JalaliradAsyncFL2019}. Asynchronous FL model is more suitable in practice, especially devices may differ in terms of hardware, network connection, and battery power, resulting in substantial heterogeneity in system parameters throughout the network~\cite{FL_challenges_review}.
The convergence bound and extension abilities of popular algorithms such as SGD  must be investigate analytically in order to give convergence guarantees for loss functions in Asynchronous FL.
Even when learning a single global model, statistical heterogeneity introduces new hurdles in assessing the convergence behavior in federated environments. 
To be specific, when data is non-i.i.d delivered among network devices, algorithms like FedAvg might diverge in practice if the chosen devices execute too many local updates. 
FedProx was recently developed to better analyze FedAvg's performance in heterogeneous contexts. 
FedProx's basic notion is that system heterogeneity and statistical heterogeneity are inextricably linked. 
FedProx modifies the FedAvg approach somewhat by allowing for partial work to be distributed among devices based on the underlying system restrictions and then using the proximal term to securely include the partial work. 
Furthermore, when applying the FL framework in RL, the widely used Q-learning or DQN (discrete actions) in solving DSS problems~\cite{RLDSA,DEQN} requires a solid theoretic proof when making convergence analysis based on FedAVG or FedProx. Therefore, policy gradient as an alternative approach is introduced in this article to update the shared model by averaging aggregated local policies.


\section{Conclusion}

This paper provides a brief overview of a Federated Learning framework to solve dynamic spectrum access problems. Particularly, Multi-agent Reinforcement Learning enabled Federated Learning (FL) is introduced. The resulting framework has the advantages of preserving user privacy and improving spectrum efficiency, offering a vision to beyond $5$G operations. We further discussed its solution in CBRS systems, accompanied by a brief review of the conceptual foundation of MARL-enabled-FL. Simulation results demonstrate that the approach can optimize the system throughput by policy gradient algorithms compared with other learning methods. The challenges and future directions of federated learning for DSA regarding in era of $5$G-Advanced are also discussed.


\bibliographystyle{IEEEtran}
\bibliography{IEEEabrv,ref}


\renewenvironment{IEEEbiography}[1]
  {\IEEEbiographynophoto{#1}}
  {\endIEEEbiographynophoto}
\section{Biographies}
\small
\vspace{0.2cm}
\noindent\textbf{Yifei Song} 
joined the ECE department at Virginia Tech as a Ph.D. student. His research interests include wireless communication, federated learning, and deep reinforcement learning.

\vspace{0.2cm}
\noindent\textbf{Hao-Hsuan Chang} 
is currently pursuing the Ph.D. degree in electrical and computer engineering at Virginia Tech. His research interests include dynamic spectrum access, echo state network, and deep reinforcement learning.

\vspace{0.2cm}
\noindent\textbf{Zhou Zhou}
is a Ph.D. candidate in the ECE department at Virginia Tech. His research interests are in the broad area of neural networks, machine intelligence and wireless communications. 

\vspace{0.2cm}
\noindent\textbf{Shashank Jere}
is currently pursuing the Ph.D. degree in Electrical and Computer Engineering at Virginia Tech. His current research interests are in the broad area of wireless communications, neural networks, machine learning/ deep learning and beyond-5G networks.

\vspace{0.2cm}
\noindent\textbf{Lingjia Liu} is a Professor and Associate Director of Wireless@Virginia Tech in the ECE department at Virginia Tech. His research interests include machine learning for wireless networks, dynamic spectrum access, emerging technologies for 5G and Beyond, and Internet of Things.
\end{document}